\documentclass[11pt,twoside,usegraphicx]{article}
\usepackage{asp2010}

\resetcounters

\begin{document}

\title{Gas Accretion onto a Supermassive Black Hole: a step to model AGN feedback}
\author{Kentaro Nagamine,$^1$ Paramita Barai,$^{1,2}$ and Daniel Proga$^{1,3}$
\affil{$^1$University of Nevada, Las Vegas, Department of Physics and Astronomy, 4505 S. Maryland Pkwy, Las Vegas, NV 89154-4002, USA}
\affil{$^2$INAF - Astronomical Observatory of Trieste, Via G.B. Tiepolo 11, I-34143 Trieste, Italy}
\affil{$^3$Princeton University Observatory, Peyton Hall, Princeton, NJ 08540, USA}}

\begin{abstract}
We study the gas accretion onto a supermassive black hole (SMBH) using the 3D SPH code GADGET-3 on scales of $0.1-200$\,pc. 
First we test our code with spherically symmetric, adiabatic Bondi accretion problem. 
We find that our simulation can reproduce the expected Bondi accretion flow very well for a limited amount of time until the effect of outer boundary starts to be visible.  We also find artificial heating of gas near the inner accretion boundary due to the artificial viscosity of SPH. 
Second, we implement radiative cooling and heating due to X-rays, and examine the impact of thermal feedback by the central X-ray source.  The accretion flow roughly follows the Bondi solution for low central X-ray luminosities, however, the flow starts to exhibit non-spherical fragmentation due to thermal instability for a certain range of central $L_X$, and a strong overall outflow develops for greater $L_X$. 
The cold gas develops filamentary structures that fall into the central SMBH, whereas the hot gas tries to escape through the channels in-between the cold filaments. 
Such fragmentation of accreting gas can assist in the formation of clouds around AGN, induce star-formation, and contribute to the observed variability of narrow-line regions. 

\end{abstract}


\section{Introduction}
Feedback from active galactic nuclei (AGN) is considered to be one of the important processes in regulating galaxy growth over cosmic time.  
Several research groups have implemented AGN feedback in cosmological hydrodynamic simulations \citep[e.g.][]{DiMatteo05, Booth09}, however they all have to assume a subgrid model for the accretion onto SMBH and feedback with uncertain parameters due to limited resolution in cosmological simulations.  

Here we take a different approach and perform small-scale simulations of gas accretion onto SMBH using the GADGET-3 SPH code \citep[originally described by][]{Springel05e} on small-scales of $r<200$\,pc.  Our long-term goal is to develop a better model of AGN feedback for large-scale cosmological simulations based on the results of small-scale BH accretion simulations. 
In this article, we report our initial results of adiabatic Bondi accretion simulations, as well as those with radiative cooling and heating by X-rays from the central SMBH  \citep{Barai11}. 
We run a large set of simulations of spherically symmetric Bondi accretion with different resolution, volume, and initial conditions.
The Bondi problem is ideal for testing any hydrodynamic code, since it has a semi-analytic solution. 


\section{Simulations of Bondi Accretion onto a SMBH}

\begin{figure}
\centering
\includegraphics[width=0.85 \linewidth]{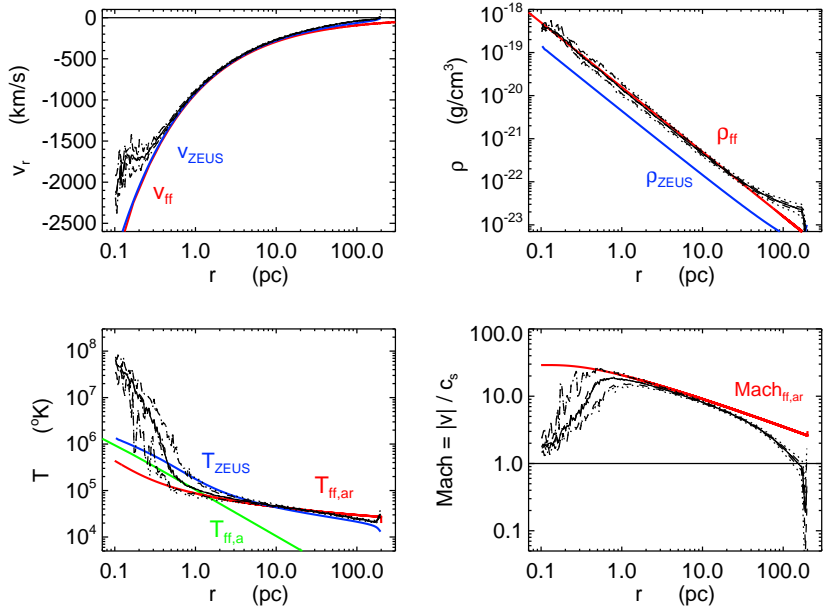}
\caption{Radial profile snapshot of gas particles in a run with X-ray heating and cooling, showing radial velocity, density, temperature and Mach number.  The solid line is the median value, the dashed lines are the 95 percentile, and the dotted lines are the min/max values.  The red curve in each panel shows the free-fall scaling, and the blue curve shows the corresponding ZEUS simulation results: $v_{\rm ZEUS}$ and $v_{\rm ff}$ are indistinguishable, $\rho_{\rm ZEUS}$ is at a constant offset from $\rho_{\rm ff}$ because of different normalization. The green curve in the bottom-left panel indicates the free-fall temperature with only adiabatic processes.  Figure taken from \citet{Barai11}.
}
\label{fig:profile}
\end{figure}

The central SMBH has a mass of $M_{\rm BH} = 10^8 M_\odot$, and the initial conditions are generated using $\gamma=1.01$, $\rho_\infty = 10^{-19}$\,g/cm$^3$, and $T_{\rm init}=T_\infty = 10^7$\,K.  The corresponding Bondi radius is $R_{\rm B} = GM_{\rm BH}/c_{s,\infty}^2 = 3.0$\,pc, the Bondi time is $t_B = R_B / c_s = 7.9\times 10^3$\,yrs, and the sonic radius is $R_s = 1.5$\,pc.   The total initial mass of the gas sphere is in the range of $10^5 - 10^7 M_\odot$ depending on the setup, and the particle counts were varied between $64^3 - 256^3$ for resolution test. 
The inner boundary radius is $r_{\rm in} = 0.1$\,pc from the central SMBH, and we regard that the particle is accreted to SMBH once it crosses $r_{\rm in}$.  The outer boundary radius was varied between $r_{\rm out} = 5 - 200$\,pc. For the full list of runs and parameters, see Table~1 of \citet{Barai11}.  

First we find that the adiabatic simulations reproduce the expected Bondi solution very well for a limited time.  
Since the gas particles near the outer boundary escape due to adiabatic expansion, after a certain time the mass accretion rate start to decrease from the Bondi rate.
This is a problem owing to the difficulty of setting a boundary condition in the SPH method, and it does not arise with the mesh codes, as one can simply fix the outer boundary with $\rho_\infty$ and $T_\infty$. 

In the second set of runs with radiative cooling and heating by X-rays, we follow the approximate treatment of \citet{Blondin94} and \citet{Proga07}, assuming that the 10 keV bremsstrahlung radiation from the central SMBH is illuminating the optically thin gas.  
In Figure~\ref{fig:profile}, we show the particle radial profile for various quantities in a run with $L_X=0.01 L_{\rm Edd}$, where $L_{\rm Edd}$ is the Eddington luminosity. 
We find that the simulation roughly follows the Bondi flow solution even with the X-ray heating and cooling, except at $r<0.5$\,pc where we see a spurious heating due to artificial viscosity (AV) of SPH particles.  We have confirmed that this effect is due to AV by turning it on and off.  With no AV, the overheating does not occur, and the gas particles overshoot the SMBH due to lack of viscosity and many are not accreted.  The mass inflow rate at $r_{\rm in}$ in this run is enhanced over the Bondi accretion rate by a factor of a few due to cooling.


\section{Non-spherical Fragmentation due to Thermal Instability, and Outflows due to X-ray Thermal Feedback}

To study the effect of X-ray thermal feedback, we gradually increase $L_X/L_{\rm Edd}$ from $5\times 10^{-5}$ to $5\times 10^{-2}$.  
In these runs, the initial condition contains 12.7 million particles for a gas sphere of $9.77\times 10^6 M_\odot$, each gas particle mass of $0.791 M_\odot$, adiabatic index $\gamma = 5/3$, $r_{\rm out}=200$\,pc, $\rho_\infty = 10^{-23}$\,g/cm$^3$, and $T_\infty=10^5$\,K.  
The Bondi radius for these parameters is 183.9\,pc, therefore most of our computational volume is within the Bondi radius and we fully resolve below the Bondi radius with minimum smoothing length of $\sim 0.1$\,pc.
In general we find that, with increasing $L_X$, the mass accretion rate at $r_{\rm in}$ decreases, and the outflow rate at $r_{\rm out}$ increases.   

When $L_X$ becomes $\ge 0.01 L_{\rm Edd}$, we start to see an interesting transition from inflow to outflow, and non-spherical fragmentation of gas into multiphase medium takes place due to thermal instability.  
The filamentary cold gas continues to flow in, and the hot gas tries to escape through the channels between cold filaments. 
Examination of the flow motion shows that the filaments get stretched, fragments, and the `clouds' merge. 
Such fragmentation of accreting gas can assist in the formation of clouds around AGN, induce star-formation, and contribute to the observed variability of narrow-line regions. 

\begin{figure}
\centering
\includegraphics[width=0.45 \linewidth]{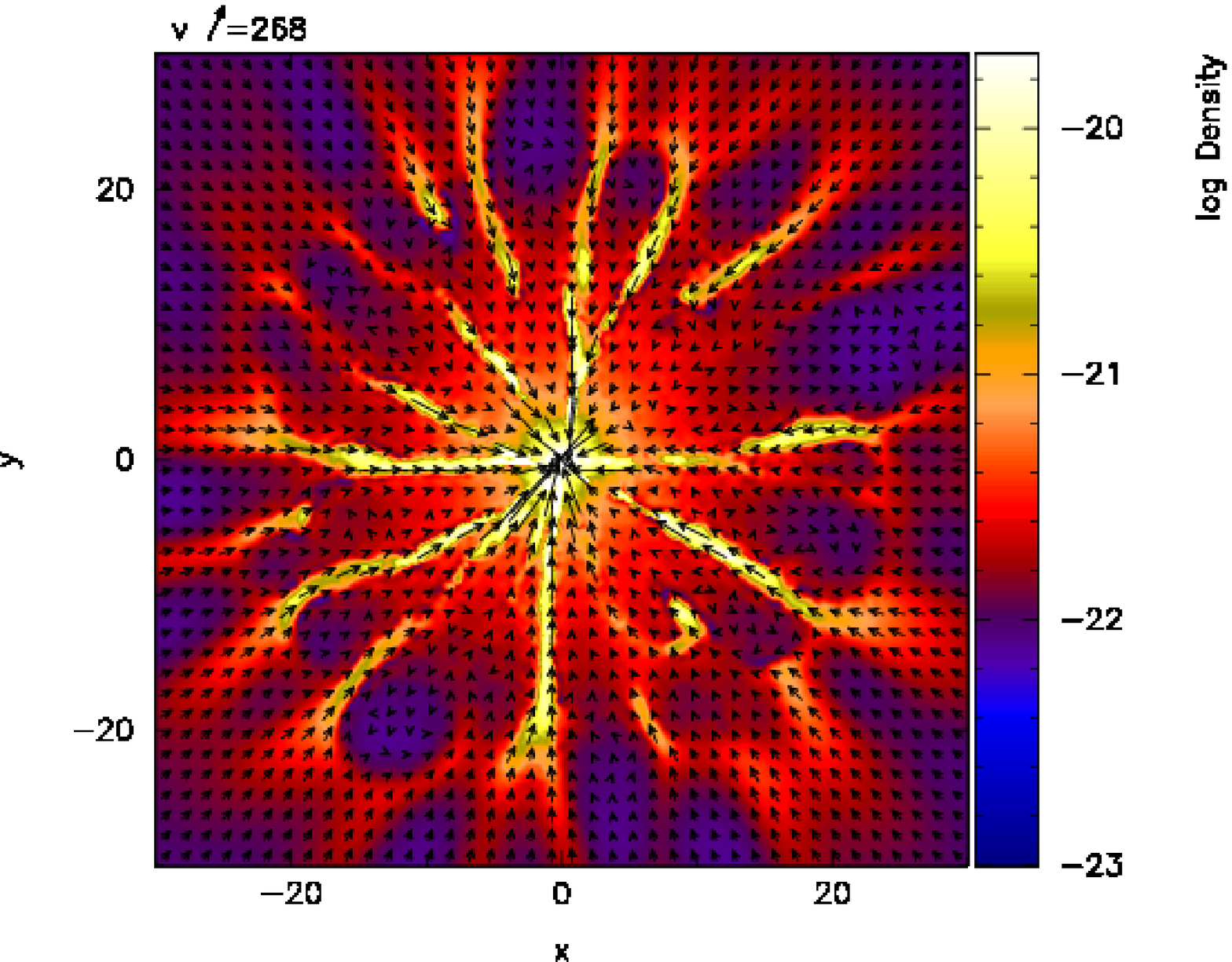}
\includegraphics[width=0.45 \linewidth]{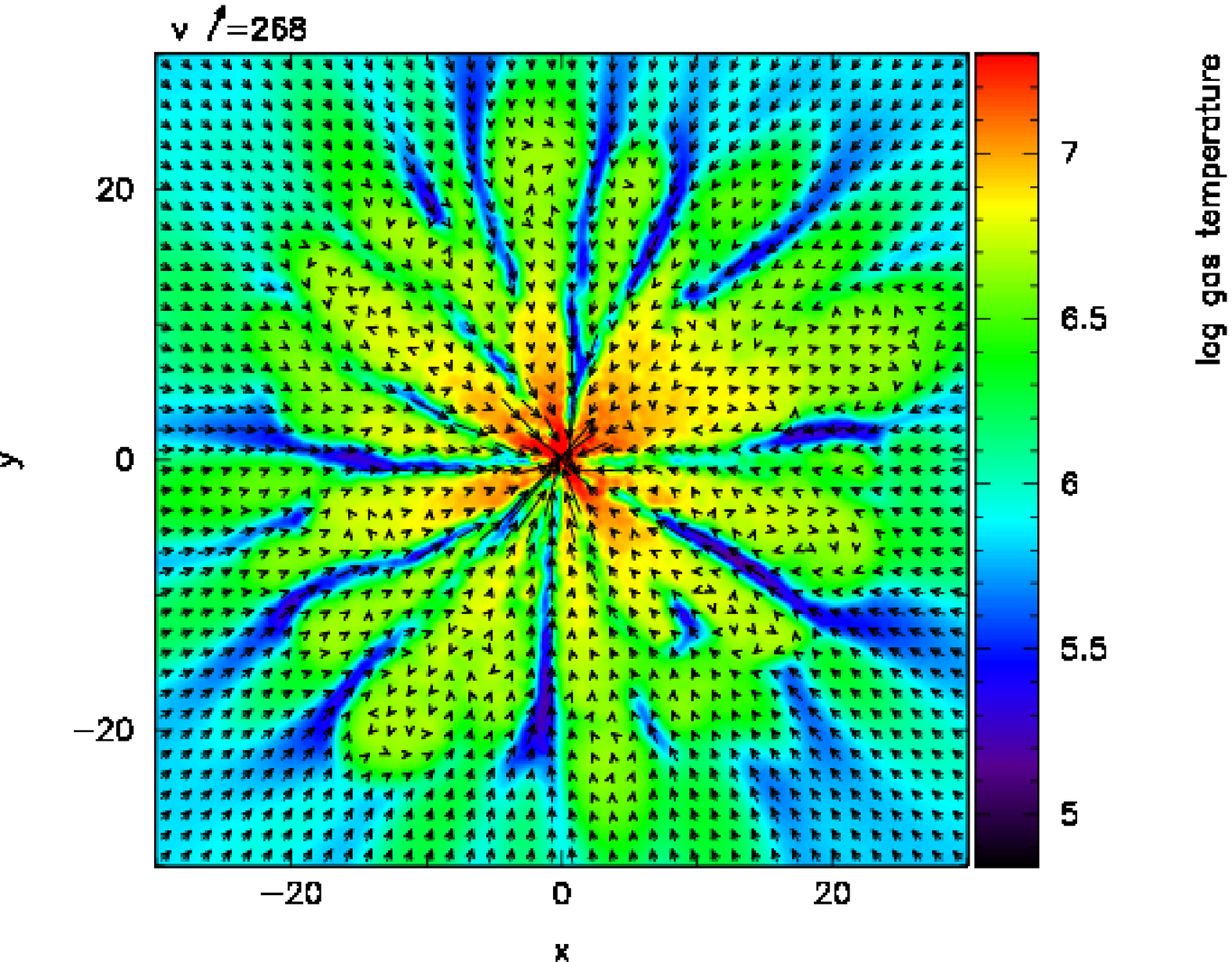}\\
\includegraphics[width=0.9 \linewidth]{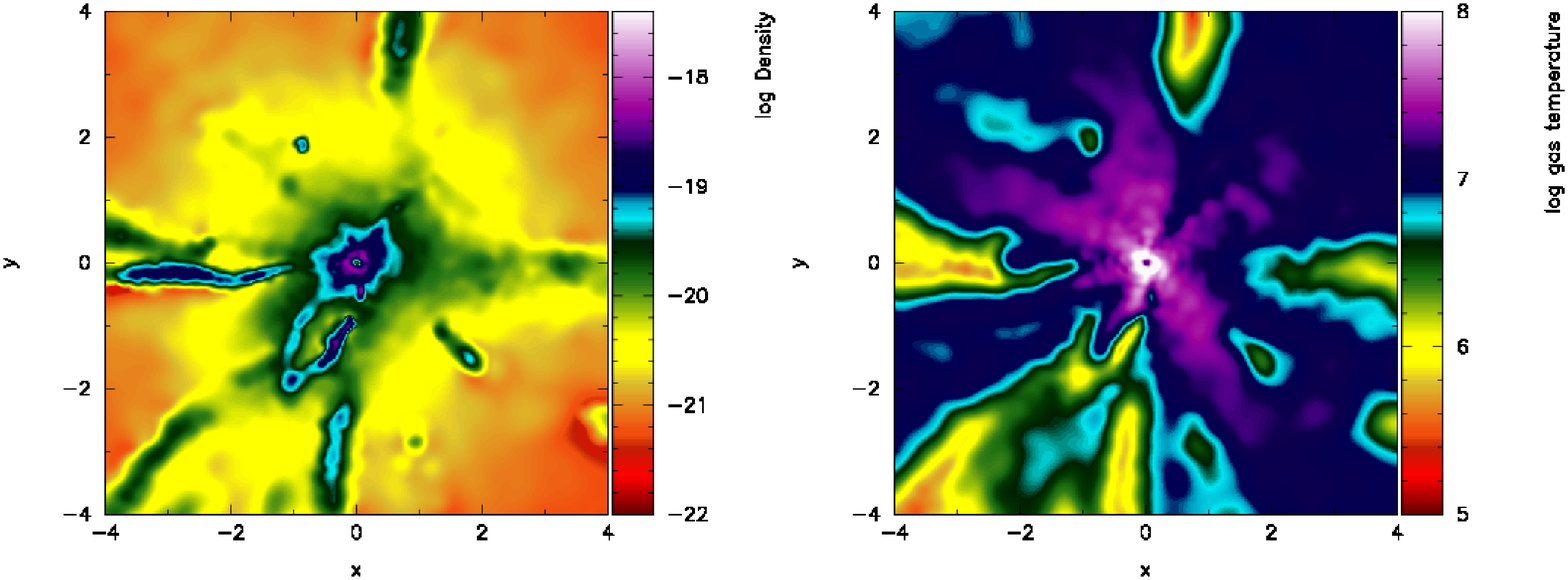}\\
\caption{Density and temperature cross-sections in a run with $L_X / L_{\rm Edd} = 0.01$.
The top two panels show the inner $\pm 30$\,pc, and the bottom two the inner $\pm 4$\,pc of the $[x - y]$ plane through $z = 0$.  
Cold, dense filamentary structure has developed due to thermal instability, which is falling into the SMBH rapidly. 
The hot gas is trying to escape, but a strong outflow has not developed yet due to low $L_X$. 
}
\label{fig:rhotemp}
\end{figure}

Figure~\ref{fig:rhotemp} shows an example of such multiphase structure in a run with $L_X = 0.01 L_{\rm Edd}$, when the outflow is still not very prominent and the hot gas is still being accreted to SMBH. 
Studies with 1D \& 2D ZEUS simulation shows that this fragmentation does not occur in a spherically symmetric Bondi flow, unless some perturbation is introduced by hand.  The SPH simulations inherently have tiny fluctuations in the density field due to its algorithmic nature, which can be amplified 	through thermal instability. 

\begin{figure}
\centering
\includegraphics[width=0.48 \linewidth]{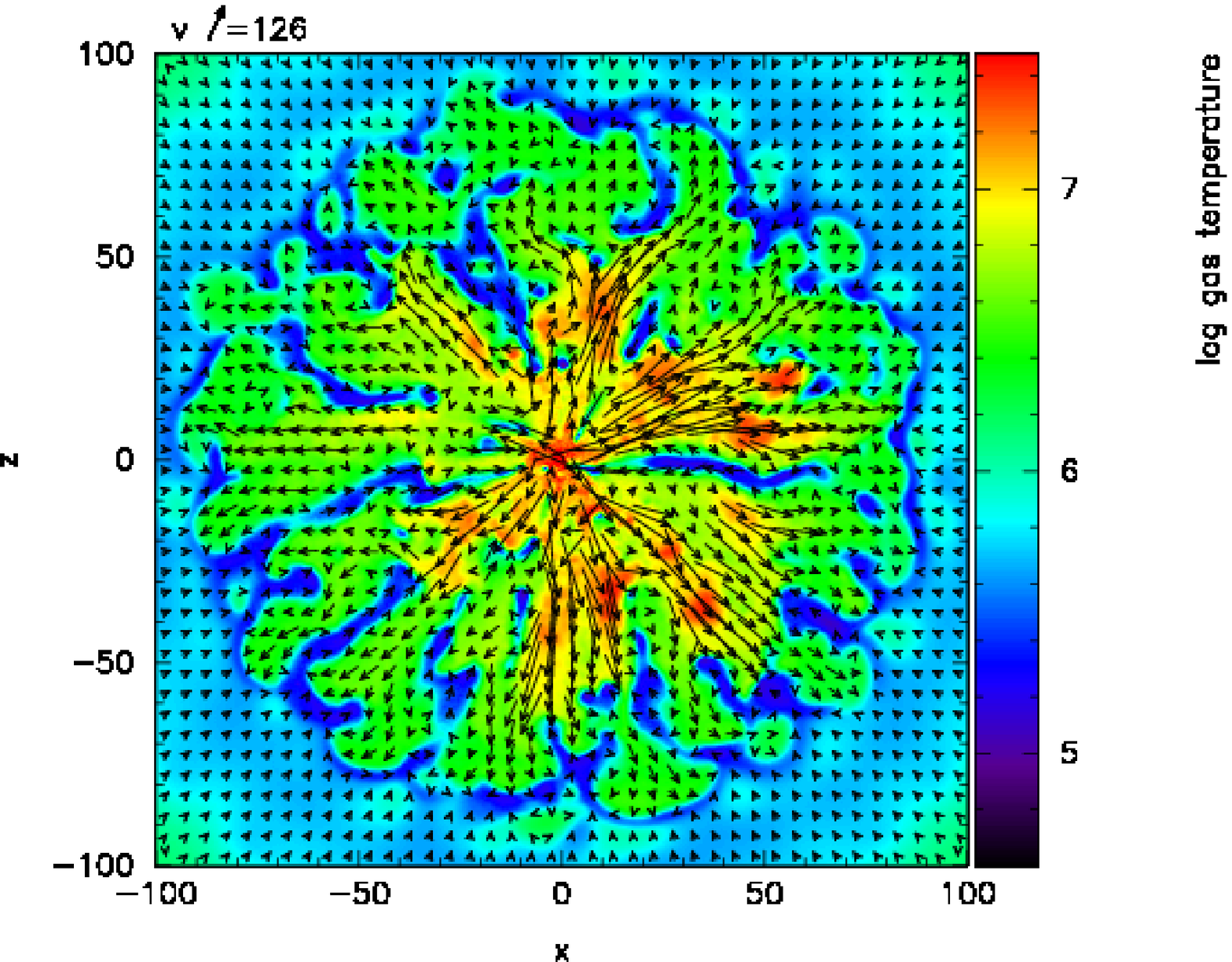}
\includegraphics[width=0.48 \linewidth]{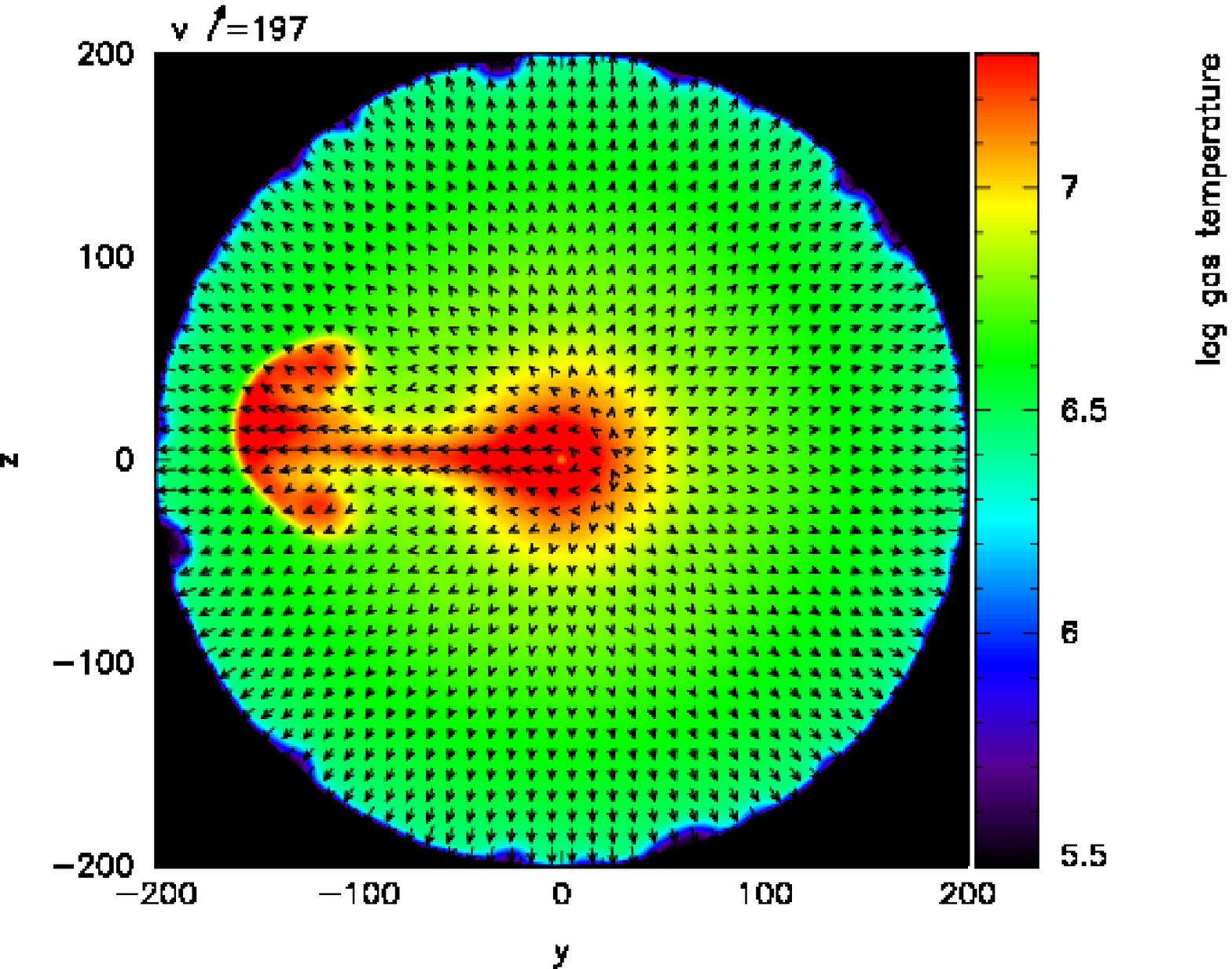}
\caption{Temperature cross-section in runs with $L_X / L_{\rm Edd} = 0.02$ (left) and 0.05 (right).
The left panel shows the inner $\pm 100$\,pc, and the right panel shows the entire computational volume of $\pm 200$\,pc. 
In the left panel, the cold, dense, filamentary structure has developed due to thermal instability, which is falling into the SMBH. 
The hot gas tries to escape through the channels in-between the cold filaments.  In the right panel, the outflow dominates over the entire computational volume, and the hot plume of gas is escaping from the center to the outer boundary. 
}
\label{fig:highLx}
\end{figure}

As $L_X$ is increased to $>$\,$0.01 L_{\rm Edd}$, the outflow starts to dominate over the inflow, and the hot gas escapes through the channels between cold filaments, as shown in the left panel of Figure~\ref{fig:highLx}.  We find that the transition from inflow to outflow occurs in-between $L_X/L_{\rm Edd} = 0.01-0.02$, but note that this transition luminosity would depend on the value of $\rho_\infty$.  In other words, the more relevant parameter for the transition is the range of photoionization parameter $\xi \propto L_X / \rho$ for the unstable branch of the $T-\xi$ equilibrium curve. 

With a high enough $L_X$ ($= 0.05 L_{\rm Edd}$), a strong outflow is produced due to strong thermal feedback, and the mass outflow rate at $r=r_{\rm out}$ increases dramatically.  A jet-like plume of hot and buoyant gas escapes from the central region to the outer boundary, as shown in the right panel of Figure~\ref{fig:highLx}. 

These results on the non-spherical flow have been reported in our second paper in series \citep{Barai12}, where we examined the $T-\xi$ equilibrium curve in detail and determine the unstable $\xi$-range.

\section{Conclusions}
We find that GADGET-3 SPH code can reproduce the spherically symmetric Bondi accretion flow properly with mainly two limitations: 1) the gas particles escape from the outer boundary due to adiabatic expansion; 2) spurious heating is observed near the inner boundary around SMBH due to the artificial viscosity of SPH. 

We also examined the impact of radiative cooling and heating due to X-rays.  The accretion flow roughly follows the Bondi solution when the central X-ray luminosity is relatively low, but the mass accretion rate at $r_{\rm in}$ is enhanced by a factor of a few over the Bondi accretion rate due to cooling. 
The simulation starts to exhibit non-spherical fragmentation due to thermal instability once $L_X$ exceeds $\simeq 0.01 L_{\rm Edd}$. 
At $L_X=0.02 L_{\rm Edd}$, the hot gas escapes in-between the cold filaments. 
When $L_X$ is further increased to $0.05 L_{\rm Edd}$, the outflow completely dominates over the inflow, and most of the gas escapes from the computational volume.  A jet-like outflow feature is also observed at this $L_X$, which selects a preferential direction for hot gas to escape rapidly. 
Such non-spherical features of accreting gas can assist in the formation of clouds around AGN, induce star-formation, and contribute to the observed variability of narrow-line regions. 

In the future, we plan to implement rotation, radiation pressure, and different initial geometry of gas distribution. 
We will measure the efficiencies of thermal, radiative, and kinetic feedback, and compare them with those measured by \citet{Kurosawa09b} and those used in the cosmological simulations.

\acknowledgements We are grateful to V. Springel for allowing us to use the GADGET-3 code. 
This work is supported in part by the NSF grant AST-0807491, National Aeronautics and Space Administration under Grant/Cooperative Agreement No. NNX08AE57A issued by the Nevada NASA EPSCoR program, and the President's Infrastructure Award from UNLV.  DP also acknowledges the UNLV sabbatical assistance.  
This research is also supported by the NSF through the TeraGrid resources provided by the Texas Advanced Computing Center.  Some numerical simulations and analyses have been performed on the UNLV Cosmology Cluster.


\begin{thebibliography}{}
\expandafter\ifx\csname natexlab\endcsname\relax\def\natexlab#1{#1}\fi
\expandafter\ifx\csname url\endcsname\relax
  \def\url#1{\texttt{#1}}\fi
\expandafter\ifx\csname urlprefix\endcsname\relax\def\urlprefix{URL }\fi
\providecommand{\eprint}[2][]{\url{#2}}

\bibitem[{{Barai} et~al.(2011{\natexlab{a}}){Barai}, {Proga}, \&
  {Nagamine}}]{Barai12}
{Barai}, P., {Proga}, D., \& {Nagamine}, K. 2011{\natexlab{a}}, ArXiv e-prints.
  \eprint{1112.5483}

\bibitem[{{Barai} et~al.(2011{\natexlab{b}}){Barai}, {Proga}, \&
  {Nagamine}}]{Barai11}
--- 2011{\natexlab{b}}, \mnras, 418, 591. \eprint{1102.3925}

\bibitem[{{Blondin}(1994)}]{Blondin94}
{Blondin}, J.~M. 1994, \apj, 435, 756

\bibitem[{{Booth} \& {Schaye}(2009)}]{Booth09}
{Booth}, C.~M., \& {Schaye}, J. 2009, \mnras, 398, 53. \eprint{0904.2572}

\bibitem[{{Di~Matteo} et~al.(2005){Di~Matteo}, Springel, \&
  Hernquist}]{DiMatteo05}
{Di~Matteo}, T., Springel, V., \& Hernquist, L. 2005, Nature, 433, 604

\bibitem[{{Kurosawa} et~al.(2009){Kurosawa}, {Proga}, \&
  {Nagamine}}]{Kurosawa09b}
{Kurosawa}, R., {Proga}, D., \& {Nagamine}, K. 2009, \apj, 707, 823.
  \eprint{0906.3739}

\bibitem[{{Proga}(2007)}]{Proga07}
{Proga}, D. 2007, \apj, 661, 693. \eprint{arXiv:astro-ph/0702582}

\bibitem[{{Springel}(2005)}]{Springel05e}
{Springel}, V. 2005, \mnras, 364, 1105. \eprint{arXiv:astro-ph/0505010}

\end{thebibliography}

\end{document}